\documentclass[conference]{IEEEtran}

\usepackage{cite}

\ifCLASSINFOpdf
  \usepackage[pdftex]{graphicx}
\else
  \usepackage[dvips]{graphicx}
\fi

\usepackage{url}

\makeatletter
\g@addto@macro{\UrlBreaks}{\UrlOrds}
\makeatother

\usepackage{xcolor}

\begin{document}
\title{Amplifying Privacy: Scaling Up Transparency Research Through Delegated Access Requests}

\author{
\IEEEauthorblockN{Hadi Asghari}
\IEEEauthorblockA{Humboldt Institute for Internet and Society\\Berlin, Germany}
\and
\IEEEauthorblockN{Thomas van Biemen}
\IEEEauthorblockA{Berenschot\\Utrecht, Netherlands}
\and
\IEEEauthorblockN{Martijn Warnier}
\IEEEauthorblockA{Delft University of Technology\\Delft, Netherlands}
}

\maketitle

\begin{abstract}
In recent years, numerous studies have used ‘data subject access requests’ in a collective manner, to tackle information asymmetries and shed light on data collection and privacy practices of organizations. While successful at increasing transparency, such studies are quite hard to conduct for the simple fact that right of access is an individual right. This means that researchers have to recruit participants and guide them through the often-cumbersome process of access. In this paper, we present an alternative method: to ask participants to delegate their right of access to the researchers. We discuss the legal grounds for doing this, the advantages it can bring to both researchers and data subjects, and present a procedural and technical design to execute it in a manner that ensures data subjects stay informed and in charge during the process. We tested our method in a pilot study in the Netherlands, and found that it creates a win-win for both the researchers and the participants. We also noted differences in how data controllers from various sectors react to such requests and discuss some remaining challenges. 
\end{abstract}

\IEEEpeerreviewmaketitle

\section{Introduction}
The \textit{right of access} to personal data is a fundamental tenet of privacy and data protection laws around the world. The right allows individuals to ask data controllers for access to any personal data that the controller holds on them, and details regarding the processing of this data. In the past decade, a growing number of researchers and activists have used this right in a systematic  manner to increase transparency in the digital economy. 
Examples include studying compliance with privacy laws\cite{ausloos_shattering_2018,mahieu_collectively_2018,norris_unaccountable_2017};  writing investigative stories\cite{duportail_i_2017,ip_who_2018}; and gathering facts for court cases\cite{cadwalladr_uk_2018,ni_loideain_end_2016}. Despite the success stories, using the right of access for transparency research is cumbersome\cite{mahieu_harnessing_2020,norris_unaccountable_2017}. This is because the right is for \textit{individuals}, meaning researchers need to recruit participants that have relations with the companies and organizations of interest to their study (in particular if they are studying long-term data collection practices and/or studying organizations that also collect data offline). Completing an access request can sometimes take months and involve several back and forth with the data controllers to overcome bureaucratic hurdles\cite{boniface_security_2019,mahieu_collectively_2018,norris_unaccountable_2017}. Individuals often do not have the time or knowledge to deal with these hurdles and achieving access can be a frustrating experience for them and the researchers. 

In this paper we outline a novel data collection method---which we are calling \textit{delegated access}---which greatly reduces the burden placed on participants of such projects. This allows researchers to scale up their transparency projects while also bolstering the rights of the data subjects. In short, the method asks study participants to delegate their access rights for a selection of data controllers to the researchers. Delegation allows the researchers to take the lead in requesting access, and handle all the hurdles that pop up along the way, such as sending reminders or arguing about a request's scope. A bonus benefit is that researchers can standardize communications with companies to a higher degree, which makes comparing and interpreting access responses easier. The method, importantly, allows participants to monitor, react, or terminate communications with the data controller as the request proceeds (should they wish). This is achieved through a  simple technical and organizational design that we shall explain later in the paper.

We tested our method in a pilot study that followed the introduction of the General Data Protection Regulation (GDPR) involving over one hundred organizations. We compared the study time with prior studies to highlight that delegated access is indeed more scalable. In addition, we found from qualitative feedback that the method empowers data subjects to gain insights that they could not have otherwise easily accessed. We observed, importantly, that delegated access works in part because the researchers are in a better position to argue with the data controllers when necessary (which is not uncommon). 

\section {Background}
\subsection{Data Processing and the Right of Access}

Privacy risks of large-scale data collection were recognized from the early days of computing and information technology\cite{westin_databanks_1972}. A key part of the solution, proposed by early thinkers in both Europe and the United States, is to guarantee the right of individuals to access and challenge the personal data held on them by data controllers. The famous American privacy scholar, Alan Westin, pointed to the need for this right in order to guarantee \textit{`due process'} as  early as 1967; and the prominent Italian data protection scholar, Stefano Rodotá  argued in 1973 that this right is fundamental for \textit{‘power reversal'}  and to shift the balance of power back to the people as organizations collect personal data\cite{mahieu_right_2019}.

The right of access has since become part of most privacy and data protection laws worldwide: it has been encoded in the OECD Privacy Principles of 1980 (\textit{“Individual Participation Principle”})\cite{organisation_for_economic_co-operation_and_development_oecd_oecd_1980}, the US Health Insurance Portability and Accountability Act of 1996 (\textit{“Right to Access Personal Health Information”})\cite{hhsgov_individuals_2020}, Hong Kong's Personal Data (Privacy) Ordinance of 1995 (\textit{“DPP 6 Access and Correction”})\cite{hong_kong_legislative_council_personal_1995}, the European Union's Data Protection Directive of 1995 (\textit{“Data Subject's Right of Access to Data”})\cite{eu_directive_1995}, the General Data Protection Regulation of 2016 (\textit{“Art 15. Right of access by the data subject”})\cite{eu_regulation_2016}, and the new California Consumer Privacy Act of 2018 (\textit{“Right to Know”})\cite{noauthor_california_2018}, just to give a few examples. Article 15a of the GDPR defines it as follows: 

\begin{quote}
``The data subject shall have the right to obtain from the controller confirmation as to whether or not personal data concerning him or her are being processed, and, where that is the case, access to the personal data and the following information:  
(a) the purposes of the processing; 
(b) the categories of personal data concerned; 
(c) the recipients or categories of recipient to whom the personal data have been or will be disclosed […]"
\end{quote}

\subsection{Using Access to Research Transparency}
Although the right of access has been established legally for a very long time, it remained veiled in obscurity from the general public for many years---until the efforts of privacy activist Max Schrems in 2011. 
After a frustrating and illuminating experience in exercising his right of access with Facebook,
he set up a tutorial to encourage others to submit access requests to Facebook, which resulted in over 40,000 requests\cite{kuchler_max_2018, noauthor_europe-v-facebookorg_nodate}. This demonstrated a public interest in the right. 

Based on this interest, digital rights groups in different countries  
have since created websites to help citizens exercise their right to access. Examples include: \textit{personaldata.io} (by Paul-Olivier Dehaye), \textit{mydatadoneright.eu} (by Bits of Freedom and partners), \textit{accessmyinfo.ca} (by Citizen Lab), \textit{accessmyinfo.hk} (by Chinese University of Hong Kong and partners), and \textit{selbstauskunft.net} (a commercial service in Germany). While these projects have made it easier for individuals to request access (with instructions, request templates, and controller contact information), they generally do not track the efficacy of these requests or the experience of the data subjects. 

A number of academic studies have attended to the question of how the right functions in practice.
One of the first major empirical studies on access rights was conducted by Norris and colleagues \cite{norris_unaccountable_2017} across ten European countries in 2013; They reported a big disparity between privacy rights in theory and practice. For instance, while trying to send access requests, researchers were unable to locate the data controller in about 20\% of cases; Among submitted requests, less than half  yielded a `positive' outcome (responses that adequately addressed the researcher's questions). These researchers identified six \textit{`discourses of denial'}---tactics deployed by data controllers to wittingly or unwittingly obstruct the exercise of access. 

Other studies have drawn similar conclusions about limited compliance with the right of access, in Europe \cite{ausloos_shattering_2018,boniface_security_2019,mahieu_collectively_2018,herrmann_obtaining_2016,spiller_experiences_2016}, as well other jurisdictions where it has been tested, for instance Canada\cite{hilts_access_2015}. Mahieu and colleagues \cite{mahieu_collectively_2018} found that among the 80\% of data controllers who eventually responded to the access requests, only half responded within the legal time limit, and one third only after a reminder. Furthermore, the majority of the responses were in some form incomplete.
Boniface and colleagues \cite{boniface_security_2019} have highlighted that the procedures used by data controllers to authenticate the data subject's identity can lead to security risks. The fact that many organizations do not live up to their legal obligations, and intentionally or unintentionally build barriers to access, disempowers data subjects \cite{mahieu_harnessing_2020}.

It is hard to judge from a single access response whether the information returned by the controller includes all the personal data they hold. 
Comparing responses to similar requests in a \textit{collective manner} can create a context to judge the quality of a response and the data practices it reveals, i.e., multiple requests to the same controller or requests sent to other controllers within the same sector. In addition to academics, NGOs and journalists have also exercised collective requests to uncover deeper insights and increase transparency. One great example is the OpenSCHUFA project \cite{openschufa_openschufa_2019}, a crowdsourcing campaign which helped show that the leading credit scoring agency in Germany was violating GDPR access provisions.

The collective exercise of the right to access, unfortunately, has major scaling limitations. As explained earlier, 
researchers (or activists) need to find  study participants that have existing relations with the data controllers of interest and are also interested in investing the time to exercise their privacy rights. 
Although some researchers\cite{bufalieri20} have resorted to first setting up accounts on various websites so they can then request access, this solution misses long-term data collection trends, and organizations that primarily collect data in `offline' environments (such as dentists or city councils). Thus depending on the transparency question, recruiting participants remains necessary. 

Participatory research has its fair share of challenges in general \cite{leavy_research_2017,kimura_citizen_2016}, and these are only intensified when it comes to the naturally private matters of access requests. Due to the accumulation of challenges in exercising access, participants often feel overwhelmed, and either cancel requests, or may even drop out of the research all together. In the OpenSCHUFA campaign, for instance,  only 4,000 out of the 30,000 requests were completed and shared with the project at the end.

\section{The Delegated Access Method}
\subsection{Method Goals}

We have developed the delegated access method for two simultaneous purposes: (i) to help reduce the barriers to access for individual data subjects, and (ii) to overcome the scaling challenge for researchers.

The method creates a win-win collaboration between data subjects and researchers: by delegating communications with data controllers to a researcher, data subjects can make use of the researcher's skill and experience in navigating the barriers to exercising access. The researchers, on the other hand, gain access to a bigger pool of data controllers to investigate. 

Although the idea of a win-win relationship between researcher and volunteer in participatory research is simple enough, it can be challenging to achieve in practice \cite{salganik_bit_2018}. Specifically in our context, since the personal data that will be transmitted is sensitive by nature, delegation demands trust among the parties and in the method itself, which we shall discuss in the next sub-sections. 

\subsection{Legal Standing}
A key question is the legality of delegating one's right of access to researchers.  

If we consider the right as it is defined in the GDPR, there is no direct mention of exercising access through third parties---even though legal representation is mentioned in other articles, e.g. Art. 35 or Art. 80. Choosing another to exercise legal actions on one's behalf is, however, well established in Western legal systems since the early days of the Roman empire \cite{eder_powers_1950}, and is enshrined in the legal codes of  many democracies. The GDPR does not limit this power.

The data protection regulatory body of the United Kingdom, the Information Commissioner's Office (ICO), explicitly mentions representation for access:
\begin{quote}
``The GDPR does not prevent an individual making a subject access request via a third party. Often, this will be a solicitor acting on behalf of a client, but it could simply be that an individual feels comfortable allowing someone else to act for them. In these cases, you need to be satisfied that the third party making the request is entitled to act on behalf of the individual, but it is the third party's responsibility to provide evidence of this entitlement. This might be a written authority to make the request or it might be a more general power of attorney."\cite{ico_right_2018}
\end{quote}

The way to authorize others to legally act on your behalf is regulated by national law.
Importantly, representation is generally not restricted to legal professions.  In the Netherlands, where the delegated access method was first used, Dutch property law specifically states that civilians have the right to declare a representative (who does not need to be a lawyer) to perform legal action on their behalf\cite{noauthor_burgerlijk_1992}. A written declaration, signed by both the subject and researcher provides sufficient proof. Comparative laws for exercising legal action through a representative can be found in other EU countries---although requirements for the proof of consent may vary\cite{busch_principles_2003}. 

\subsection {Inception Phase}
\begin{figure}[!t]
\centering
\includegraphics[width=2.5in]{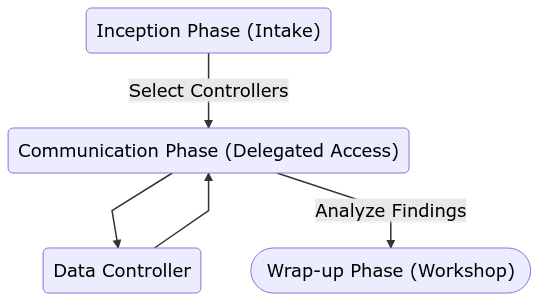}
\caption{Delegated Access Workflow}
\label{fig_process}
\end{figure}

To balance the goals of empowerment and scalability we propose to organize delegated access in three distinct phases, highlighted in Fig. \ref{fig_process}. The method starts with the \textit{inception phase}, in which the researcher recruits and informs a new group of participants. In order to be included in this phase,  recruits need to have data relations with organizations of interest for the research, and ideally be genuinely interested in the results of the access requests that will be sent on their behalf.  

Because it is important that participants understand the risks associated with the research and trust the researcher in handling these risks, we recommend that a \textit{one-on-one meeting} is scheduled for each participant. In this meeting, participants are informed of the research goals and process; 
They are then asked to write down an extensive list of companies and organizations that they expect might have their personal data. 
The participant also indicates for each organization how interested they are in their reply, if the data received from this organization may indeed be used for research and if the communications with this organizations may be performed by the researcher on their behalf. 
After these agreements, the participant is asked to sign the research consent and delegation form. The participant also shares a secure copy of his proof of identity with the researcher (which is necessary for submitting access requests). 

When all participants have finished this inception phase, the researcher will decide which of the recorded organizations to contact on whose behalf. In this choice, the researcher 
takes into account that requests may require time investments of the participant, and should seek to access data for no more than about ten organizations per participant. 
The final list of organizations to be contacted on their behalf is then communicated to all participants, giving them a final say 
before starting communications in case they have changed their mind. Since the participant might not necessarily know what data a controller holds on them, they should be free to change their mind about the inclusion or exclusion of an organization from the research as they receive access. 

\textit{The primary ethical consideration in this phase is to guarantee informed consent.} While informed consent is a general requirement in research involving human subjects, the sensitivity of personal data involved in data access makes it even more critical. 
We designed for this requirement by:
\begin{itemize}
    \item Meeting all participants in person once, to ensure they understand the research goal, process, and risks sufficiently;
    \item Obtaining consent for access to each organization separately (instead of asking for a `blanket' consent); 
    \item Notifying participants of their right to 
    change their consent for any organization at any point. 
\end{itemize}

\textit{Another consideration is the load on the organizations that will be targets of the research}. Although legally organizations should be able to respond to access requests by all their data subjects, this is unfortunately not yet the reality in many cases. We thus propose limiting the number of requests sent to each data controller, especially when dealing with smaller organizations that may be excessively burdened. 

\subsection{Communication Phase}
The \textit{communication phase} is the heart of the method. All access requests are submitted to data controllers using a standardized request letter\footnote{We adapted request examples provided by several  European Data Protection Authorities.}. This request must be accompanied by a copy of the signed communication consent form and the proof of identities of the researcher and participant. It is sent using the communication method that is described in the organization's privacy statement (e.g., email, web form, or physical mail, with email being our preferred option). 

The communication phase continues with both participants and researchers responding to organizations' replies to the request. 
As previously discussed, we found it vital to allow participants to remain fully informed of the communication, and if they desire, to take over. For this purpose, 
we created our own digital system ( Fig. \ref{fig_system}). The key design insight is to \textit{centralize communications using a custom email platform}\footnote{We used the open source Roundcube webmail program (https://roundcube.net) and Postfix 
mail server (https://postfix.org).}\textit{and domain}, as follows:

\begin{figure}[!t]
\centering
\includegraphics[width=2.5in]{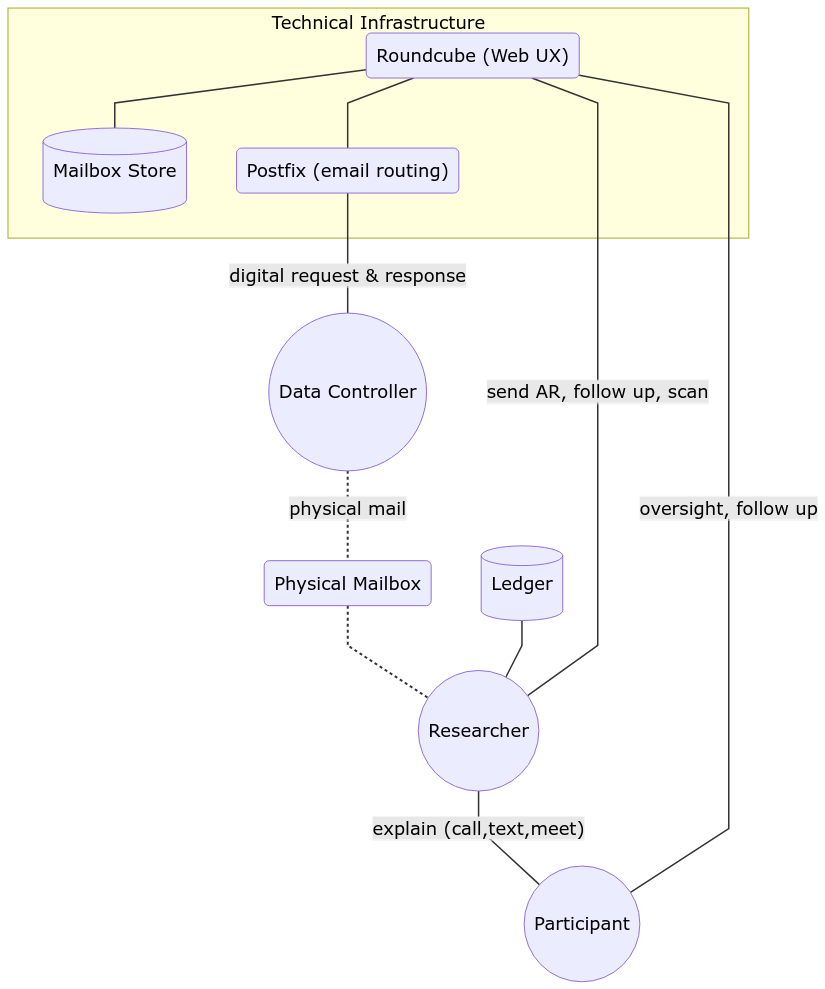}
\caption{Delegated Access System Design}
\label{fig_system}
\end{figure}

\begin{itemize}
    \item  Each study participant receives their own mailbox in the form of $<lastname@projectdomain>$ to which they have access at all times;
    \item  The researcher can send emails on behalf of participants to an organization; 
    a blind copy\footnote{Implemented using a combination of Roundcube and Postfix features.} is always delivered to the participant inbox for their information;
    \item  When an organization responds, the email is delivered to the corresponding participant's mailbox; if the participant has agreed to share all responses of this controller with the researcher, then a blind copy is also delivered to the researcher's inbox\footnote{If the participant has chosen not to automatically share responses, they may still manually forward sections
of an email to the researcher.}; 
    \item  Participants can respond to any email directly, although the idea is to let the researchers handle correspondence. 
\end{itemize}

Some data controllers require or respond only to physical letters.\footnote{This may be specific for certain sectors or countries where specific regulations differs with the GDPR on this point. See the illustrative case for more information on these occurrences.} For this reason, a central (physical) mailing address is added to the signature of every request. Digital scans of all incoming or outgoing letters are also emailed to the participants.\footnote{Letters received on a participant’s behalf can be opened and answered if that participant has authorized it; otherwise they remain unopened. When authorized, we still asked participants for permission before opening each individual letter (by text), because of the stricter expectation of privacy for physical mail in the Netherlands.} Similarly, when other communication methods are used with a data controller (such as phone calls, personal conversation, or webforms), the researchers share the contents with the participant digitally, so a central log is maintained. 

\textit{The key ethical consideration in this phase is data privacy and protection}. In our case, all processing of personal data was handled on university infrastructure (not the public cloud). Our servers were administrated professionally with appropriate patching, firewalls, HTTPS access, backups, etc. 

\textit{We already touched upon transparency towards the participants}. A related ethical dilemma is whether to reveal or conceal from the data controllers that we are conducting academic research.  Our approach was to not explicitly mention this fact in our correspondence, but also not to hide it. We provided information about our project on the mail domain, which some curious controllers visited. Our motivation for this choice, other than the virtue of transparency, was that since delegated access is not (yet) a mainstream method, some data controllers might be understandably cautious in sharing personal data with third parties. 

\subsection{Wrap-Up Phase}
When data collection has reached a point where organizations have received appropriate time to correctly respond to the   requests, the researcher stops the collection phase. 
All participants are then invited to a \textit{workshop} where the aggregated project insights are shared with them. During this workshop, participants are also asked about their experience. 

\textit{A final ethical consideration is data minimization with regards to storing the access responses}. We suggest anonymizing and/or pseudonymizing the responses as soon as is possible after the analysis phase.

\section{Illustrative case }
We evaluated the delegated access method in a pilot study conducted between June to October 2018, immediately after the GDPR became enforceable. 
In this section, we shall briefly describe the study setup and the key results and observations. The full study is openly available as a master thesis~\cite{biemen2018}.

\begin{table*}[t]
\caption{Statistics per access request comparing the delegated access pilot with other large-scale studies of access rights.}
\label{table_results}
\centering
\begin{tabular}{|c|c|c|c|c|c|c|c|}
\hline
Study & Controllers & Researchers & Participants &  Access Duration & Time/AR Researcher & Time/AR Participant & Response  \\
\hline
Norris  et al.\cite{norris_unaccountable_2017} & 183  & 19 & - &  ±6 months  &  Many hours & - & 80\% \\
Ausloos \& Dewitte\cite{ausloos_shattering_2018} & 60 & 2  &  3 &  4 months & 1 hour & 1.5 hours & 74\% \\
Mahieu et al.\cite{mahieu_collectively_2018} & 106 & 3 & 7 & 4 months & 1 hour & 1 hour & 83\% \\
Our Pilot  & 116 & 1 & 10 & 3 months & 1.5 hours & 20 minutes & 81\% \\
\hline
\end{tabular}
\end{table*}

\subsection{Pilot Study Setup}
Following the guidelines presented in Section III, university ethics research approval was first obtained for the study. Next, a call for participation was distributed among the family, friends, and friends of friends of the pilot study's lead researcher.  Over 37 participants signed up, and the first ten participants were selected for the pilot. 

Following our protocol, an intake interview of approximately one hour was conducted with each participant. During the intake interview, 
each participant consented to delegating their right of access to a set of organizations.  All participants volunteered enough organizations to allow choosing a dozen interesting ones. 
Perhaps not surprisingly, participants closer to the researcher volunteered a larger number of potential organizations for the research. 

To prove the participant's identity, a secure copy was made of their identity card using an app provided by the Dutch government specifically for this goal: \textit{KopieID}\cite{kopieid}.

At the end of the inception phase, around 120 organizations were chosen for requesting access. Since all participants were Dutch inhabitants, the scope of the research was  limited to organizations that control data of Dutch residents.\footnote{The controllers  included a diverse set of companies and organizations. Approximately 67\% were local Dutch entities, 23\% were multinationals headquartered in the EU, and 10\% were multinationals headquartered outside of the EU. Approximately 55\% of the controllers were large entities (with over 250 employees) and 25\% were small (with under 50 employees). For more details see  \cite{biemen2018}. } All organizations where contacted using a letter asking for a copy of the personal data that is processed, and information concerning the source of the data, processing goals, organizations with whom the data is shared, and retention periods, via the contact method suggested in their privacy policy. 
 
\subsection{Results} 
Overall, 81\% of the organizations in the sample responded to requests for access. This response rate is comparable with three other large-scale studies of access rights (see Table \ref{table_results}). This suggests that delegation performs similarly in this regard to studies involving only researchers \cite{norris_unaccountable_2017} or having participants work under the guidance of researchers \cite{ausloos_shattering_2018,mahieu_collectively_2018}. The overall compliance rate is lower (at 51\%) if we consider the legal time limit and quality of the access responses. This rate is higher than the earlier studies which had an average compliance rate of 33\%. This increase may be due to either the positive influence of the GDPR or the use of delegation---more research is needed to parse out these factors. 

\textit{The key improvement lies in the scalability aspect}. Using the delegated method, one researcher can submit more than 100 requests on behalf of 10 participants, which is about three times higher than earlier studies\footnote{Times for other studies are either derived from the publications or in correspondence with the authors.}. 
The average time investment per request for the researcher is 50\% more than the other studies involving participants; the time investment for participants, on the other hand, drops to about a third because of the researcher's help; an overall net time 
gain\footnote{The researcher time may be further reduced in the future as delegation is more widely used, since many controllers challenged its legality at first.}.
The increased scalability is in addition to the increased empirical rigor that comes from handling access requests in a standardized manner (which improves the comparability of responses). 

\textit{Overall study participants expressed positive attitudes towards the method}. During the final workshop, participants expressed that they did not think they could have gained the same insights without the help of the researchers, given the number of back and forth involved with some organizations.\footnote{An improvement suggested by some participants is to send them only a weekly digest with the only most important communications.}

\subsection{Discussion}
In applying the novel method, we recognized each of the discourses of denial that were identified by earlier studies. And although the introduction of the GDPR seemed to help in overcoming some of these hurdles, the \textit{out of court} discourse~\cite{norris_unaccountable_2017} was very prevalent, as many data controllers did not easily acknowledge the lawfulness of the delegated access request. Often, references to the interpretation of the relevant GDPR rights by multiple DPOs and direct quotes of the Dutch Civil Code were necessary to convince data controllers of the legality of delegation.

Many organizations requested \textit{extra proof of identity} from the data subjects, such as an email account or telephone number they had on file, in addition to signed consent letters and the already provided proofs of identity.\footnote{According to the Dutch DPA, data controllers can ask respondents to identify themselves by sharing a copy of their identity card but should strive to opt for less intrusive methods. We were asked for additional proof \textit{on top of} the shared identity cards of both the researcher and the participant.} 
These requests were handled on a per case basis in consultation with the participant. Governmental organizations showed a distaste for electronic communication methods, which added to the burden of access. There was an additional challenge related to the Dutch healthcare sector, where contradictory health-specific regulations barred the full sharing of personal data as detailed by the GDPR.\footnote{This contradiction may be a symptom of two clashing public values, confidentiality versus transparency of data processing.} 

Dealing with these challenges highlighted the initially underrated strength of the delegated access method: legal and bureaucratic hurdles could be dealt with by the researchers with expertise and caused less personal tension. In this manner, \textit{delegation reduces the power asymmetry} that  exists.  

\textit{Contrary to our expectations, smaller organizations on average handled access requests better}. This may be because smaller organizations are more prone to invest time into a personal answer, a finding that is also supported by further qualitative analysis. 

Another unexpected situation concerns \textit{the personal relationships of data subjects with certain data controllers}. As explained, using delegation shields data subjects from personal pressures in the process, which not uncommonly leads them to withdraw their access request. In a few occasions, however, data subjects were still approached directly by the organization and pleaded with to withdraw their request (because, for instance, it would take the organization too much work to respond). One easy improvement lies in warning participants that such things may happen, and to inform them that they can politely dismiss such pleas. Controllers could also be warned of implications of discouraging people to practice their legal rights. 

While delegated access enhances the scalability of research that uses access requests, it can be questioned whether it is truly scalable, given that the researcher still needs to sit down with each participant and spend time communicating with each data controller. This is especially apparent if one considers alternatives, such as asking volunteers to download their data from websites and donating it to the research project. We believe that these methods are complimentary and suitable for different transparency questions. For instance, smaller and less technically capable data controllers often do not offer download tools. Additionally, the courts have ruled that the scope of the data covered by download tools is less\footnote{See the part about machine readable files in the recent Dutch court case regarding the access rights of Uber  drivers~\cite{uber_2021}} than what the right of access covers. 

One possible future research topic is the introduction of gaming elements (e.g., as described by \cite{salganik_bit_2018}) to find the optimal balance between researcher and participant effort, and to make sure participants remain motivated throughout the project.

\section{Conclusion}
Delegated access aims to allow researchers to create a larger and more relevant dataset of access responses for their research, while also delivering genuine insights to data subjects. The method, as discussed in this paper and evaluated in a pilot study, succeeded in both aspects. Although this paper focuses on using delegated access under the GDPR, we believe it can also be applied in other jurisdictions, since the legal basis (access rights and delegation) holds in many countries. The pilot also identified a number of challenges—for instance with controllers being uncertain of the legality of delegation. As with the right of access itself, the delegated access method will provide the most insights when its application becomes common for researchers, data subjects, and controllers.

\section*{Acknowledgment}

The authors wish to thank all the pilot study participants, as well as (in alphabetical order) Jef Ausloos, Paul-Olivier Dehaye,  Karolina Iwańska, René Mahieu, Joris van Hoboken, Frederik Zuiderveen Borgesius, plus the anonymous reviewers for their contributions and feedback to this research. This research received no specific grant from any funding agency in the public, commercial, or not-for-profit sectors.

\bibliographystyle{IEEEtran}
\bibliography{IEEEabrv,./delegated.bib}

\begin{thebibliography}{10}
\providecommand{\url}[1]{#1}
\csname url@samestyle\endcsname
\providecommand{\newblock}{\relax}
\providecommand{\bibinfo}[2]{#2}
\providecommand{\BIBentrySTDinterwordspacing}{\spaceskip=0pt\relax}
\providecommand{\BIBentryALTinterwordstretchfactor}{4}
\providecommand{\BIBentryALTinterwordspacing}{\spaceskip=\fontdimen2\font plus
\BIBentryALTinterwordstretchfactor\fontdimen3\font minus
  \fontdimen4\font\relax}
\providecommand{\BIBforeignlanguage}[2]{{%
\expandafter\ifx\csname l@#1\endcsname\relax
\typeout{** WARNING: IEEEtran.bst: No hyphenation pattern has been}%
\typeout{** loaded for the language `#1'. Using the pattern for}%
\typeout{** the default language instead.}%
\else
\language=\csname l@#1\endcsname
\fi
#2}}
\providecommand{\BIBdecl}{\relax}
\BIBdecl

\bibitem{ausloos_shattering_2018}
\BIBentryALTinterwordspacing
J.~Ausloos and P.~Dewitte, ``Shattering one-way mirrors – data subject access
  rights in practice,'' vol.~8, no.~1, pp. 4--28, 2018. [Online]. Available:
  \url{https://academic.oup.com/idpl/article/8/1/4/4922871}
\BIBentrySTDinterwordspacing

\bibitem{mahieu_collectively_2018}
\BIBentryALTinterwordspacing
R.~L. Mahieu, H.~Asghari, and M.~J. Van~Eeten, ``Collectively exercising the
  right of access: individual effort, societal effect,'' vol.~7, no.~3, pp.
  1--22, 2018. [Online]. Available: \url{https://doi.org/10.14763/2018.3.927}
\BIBentrySTDinterwordspacing

\bibitem{norris_unaccountable_2017}
\BIBentryALTinterwordspacing
C.~Norris, P.~De~Hert, X.~L'Hoiry, and A.~Galetta, Eds., \emph{The
  Unaccountable State of Surveillance - Exercising Access Rights in Europe},
  ser. Law, Governance and Technology Series.\hskip 1em plus 0.5em minus
  0.4em\relax Springer International Publishing, 2017, no.~34. [Online].
  Available: \url{http://www.springer.com/us/book/9783319475714}
\BIBentrySTDinterwordspacing

\bibitem{duportail_i_2017}
\BIBentryALTinterwordspacing
J.~Duportail, ``I asked tinder for my data. it sent me 800 pages of my deepest,
  darkest secrets,'' 2017. [Online]. Available:
  \url{http://www.theguardian.com/technology/2017/sep/26/tinder-personal-data-dating-app-messages-hacked-sold}
\BIBentrySTDinterwordspacing

\bibitem{ip_who_2018}
\BIBentryALTinterwordspacing
C.~Ip, ``Who controls your data?'' 2018. [Online]. Available:
  \url{https://www.engadget.com/2018-09-04-who-controls-your-data.html}
\BIBentrySTDinterwordspacing

\bibitem{cadwalladr_uk_2018}
\BIBentryALTinterwordspacing
C.~Cadwalladr, ``{UK} regulator orders cambridge analytica to release data on
  {US} voter,'' 2018-05-05. [Online]. Available:
  \url{https://www.theguardian.com/uk-news/2018/may/05/cambridge-analytica-uk-regulator-release-data-us-voter-david-carroll}
\BIBentrySTDinterwordspacing

\bibitem{ni_loideain_end_2016}
\BIBentryALTinterwordspacing
N.~Ni~Loideain, ``The end of safe harbor: Implications for {EU} digital privacy
  and data protection law,'' vol.~19, no.~8, pp. 1--14, 2016. [Online].
  Available: \url{https://ssrn.com/abstract=2734698}
\BIBentrySTDinterwordspacing

\bibitem{mahieu_harnessing_2020}
\BIBentryALTinterwordspacing
R.~L. Mahieu and J.~Ausloos, ``Harnessing the collective potential of {GDPR}
  access rights: towards an ecology of transparency,'' 2020-07-06, library
  Catalog: policyreview.info. [Online]. Available:
  \url{https://policyreview.info/articles/news/harnessing-collective-potential-gdpr-access-rights-towards-ecology-transparency/1487}
\BIBentrySTDinterwordspacing

\bibitem{boniface_security_2019}
\BIBentryALTinterwordspacing
C.~Boniface, I.~Fouad, N.~Bielova, C.~Lauradoux, and C.~Santos, ``Security
  analysis of subject access request procedures how to authenticate data
  subjects safely when they request for their data,'' 2019. [Online].
  Available:
  \url{https://link.springer.com/chapter/10.1007/978-3-030-21752-5_12}
\BIBentrySTDinterwordspacing

\bibitem{westin_databanks_1972}
A.~F. Westin and M.~A. Baker, \emph{Databanks in a Free Society}.\hskip 1em
  plus 0.5em minus 0.4em\relax Quadrangle Books, 1972.

\bibitem{mahieu_right_2019}
R.~Mahieu, ``The right of access: A genealogy,'' 2019, {TILTING}2019:
  Regulating a World in Transition.

\bibitem{organisation_for_economic_co-operation_and_development_oecd_oecd_1980}
\BIBentryALTinterwordspacing
{Organisation for Economic Co-operation and Development (OECD}, ``{OECD}
  guidelines on the protection of privacy and transborder flows of personal
  data,'' 1980. [Online]. Available:
  \url{https://www.oecd.org/digital/ieconomy/privacy-guidelines.htm}
\BIBentrySTDinterwordspacing

\bibitem{hhsgov_individuals_2020}
\BIBentryALTinterwordspacing
{HHS.gov} and {U.S. Department of Health \& Human Services}. (2020)
  Individuals’ right under {HIPAA} to access their health information 45
  {CFR} § 164.524. [Online]. Available:
  \url{https://www.hhs.gov/hipaa/for-professionals/privacy/guidance/access/index.html}
\BIBentrySTDinterwordspacing

\bibitem{hong_kong_legislative_council_personal_1995}
\BIBentryALTinterwordspacing
{Hong Kong Legislative Council}, ``The personal data (privacy) ordinance (cap.
  486),'' 1995. [Online]. Available:
  \url{https://www.pcpd.org.hk/english/data_privacy_law /
  ordinance_at_a_Glance/ordinance.html}
\BIBentrySTDinterwordspacing

\bibitem{eu_directive_1995}
\BIBentryALTinterwordspacing
``Directive 95/46/{EC} of the european parliament and of the council of 24
  october 1995 on the protection of individuals with regard to the processing
  of personal data and on the free movement of such data,'' 1995. [Online].
  Available:
  \url{http://eur-lex.europa.eu/legal-content/EN/TXT/HTML/?uri=CELEX:31995L0046&from=EN}
\BIBentrySTDinterwordspacing

\bibitem{eu_regulation_2016}
\BIBentryALTinterwordspacing
``Regulation ({EU}) 2016/ 679 of the european parliament and of the council of
  27 april 2016 - on the protection of natural persons with regard to the
  processing of personal data and on the free movement of such data, and
  repealing directive 95/ 46/ {EC} (general data protection regulation) {OJ} l
  119/1,'' 2016. [Online]. Available:
  \url{http://ec.europa.eu/justice/data-protection/reform/files/regulation_oj_en.pdf}
\BIBentrySTDinterwordspacing

\bibitem{noauthor_california_2018}
\BIBentryALTinterwordspacing
``California consumer privacy act of 2018 ({CCPA}),'' 2018. [Online].
  Available: \url{https://www.oag.ca.gov/privacy/ccpa}
\BIBentrySTDinterwordspacing

\bibitem{kuchler_max_2018}
\BIBentryALTinterwordspacing
H.~Kuchler, ``Max schrems: the man who took on facebook — and won,'' 2018,
  library Catalog: www.ft.com. [Online]. Available:
  \url{https://www.ft.com/content/86d1ce50-3799-11e8-8eee-e06bde01c544}
\BIBentrySTDinterwordspacing

\bibitem{noauthor_europe-v-facebookorg_nodate}
\BIBentryALTinterwordspacing
europe-v-facebook.org. Library Catalog: www.europe-v-facebook.org. [Online].
  Available: \url{http://www.europe-v-facebook.org}
\BIBentrySTDinterwordspacing

\bibitem{herrmann_obtaining_2016}
\BIBentryALTinterwordspacing
D.~Herrmann and J.~Lindemann, ``Obtaining personal data and asking for erasure:
  Do app vendors and website owners honour your privacy rights?'' 2016.
  [Online]. Available: \url{http://arxiv.org/abs/1602.01804}
\BIBentrySTDinterwordspacing

\bibitem{spiller_experiences_2016}
\BIBentryALTinterwordspacing
K.~Spiller, ``Experiences of accessing {CCTV} data: The urban topologies of
  subject access requests,'' vol.~53, no.~13, pp. 2885--2900, 2016, publisher:
  {SAGE} Publications Ltd. [Online]. Available:
  \url{https://doi.org/10.1177/0042098015597640}
\BIBentrySTDinterwordspacing

\bibitem{hilts_access_2015}
\BIBentryALTinterwordspacing
A.~Hilts and C.~Parsons, ``Access my info: An application that helps people
  create legal requests for their personal information,'' in \emph{Hot topics
  in Privacy Enhancing Technologies Symposium ({HotPETS} 2015)}, 2015.
  [Online]. Available:
  \url{https://www.petsymposium.org/2015/papers/hilts-ami-hotpets2015.pdf}
\BIBentrySTDinterwordspacing

\bibitem{openschufa_openschufa_2019}
\BIBentryALTinterwordspacing
{OpenSCHUFA}. (2019) {OpenSCHUFA}: The campaign is over, the problems remain.
  Library Catalog: openschufa.de. [Online]. Available:
  \url{https://openschufa.de/english/}
\BIBentrySTDinterwordspacing

\bibitem{bufalieri20}
L.~{Bufalieri}, M.~L. {Morgia}, A.~{Mei}, and J.~{Stefa}, ``Gdpr: When the
  right to access personal data becomes a threat,'' in \emph{2020 IEEE
  International Conference on Web Services (ICWS)}, 2020, pp. 75--83.

\bibitem{leavy_research_2017}
P.~Leavy, \emph{Research Design: Quantitative, Qualitative, Mixed Methods,
  Arts-Based, and Community-Based Participatory Research Approaches},
  1st~ed.\hskip 1em plus 0.5em minus 0.4em\relax The Guilford Press, 2017.

\bibitem{kimura_citizen_2016}
\BIBentryALTinterwordspacing
A.~H. Kimura and A.~Kinchy, ``Citizen science: Probing the virtues and contexts
  of participatory research,'' vol.~2, no.~0, pp. 331--361, 2016, number: 0.
  [Online]. Available:
  \url{https://estsjournal.org/index.php/ests/article/view/99}
\BIBentrySTDinterwordspacing

\bibitem{salganik_bit_2018}
\BIBentryALTinterwordspacing
M.~J. Salganik, \emph{Bit By Bit: Social Research in the Digital Age}.\hskip
  1em plus 0.5em minus 0.4em\relax Princeton University Press, 2018. [Online].
  Available: \url{http://www.bitbybitbook.com/}
\BIBentrySTDinterwordspacing

\bibitem{eder_powers_1950}
\BIBentryALTinterwordspacing
P.~J. Eder, ``Powers of attorney in international practice,'' vol.~98, no.~6,
  pp. 840--863, 1950, publisher: {JSTOR}. [Online]. Available:
  \url{https://scholarship.law.upenn.edu/cgi/viewcontent.cgi?article=8254&context=penn_law_review}
\BIBentrySTDinterwordspacing

\bibitem{ico_right_2018}
\BIBentryALTinterwordspacing
{ICO}. (2018) Right of access. [Online]. Available:
  \url{https://ico.org.uk/for-organisations/guide-to-data-protection/guide-to-the-general-data-protection-regulation-gdpr/individual-rights/right-of-access/}
\BIBentrySTDinterwordspacing

\bibitem{noauthor_burgerlijk_1992}
\BIBentryALTinterwordspacing
\emph{Burgerlijk Wetboek Boek 3}.\hskip 1em plus 0.5em minus 0.4em\relax
  Ministerie van Binnenlandse Zaken en Koninkrijksrelaties, 1992, last
  Modified: 2020-07-21 Library Catalog: wetten.overheid.nl. [Online].
  Available: \url{https://wetten.overheid.nl/BWBR0005291/2017-09-01}
\BIBentrySTDinterwordspacing

\bibitem{busch_principles_2003}
\BIBentryALTinterwordspacing
D.~Busch, E.~H.~Hondius, H.~J.~Van~Kooten, H.~N.~Schelhaas, and W.~M.~Schrama,
  ``The principles of european contract law and dutch law, a commentary,''
  vol.~55, no.~3, pp. 706--708, 2003, publisher: Persée - Portail des revues
  scientifiques en {SHS}. [Online]. Available:
  \url{https://www.persee.fr/doc/ridc_0035-3337_2003_num_55_3_18983}
\BIBentrySTDinterwordspacing

\bibitem{biemen2018}
\BIBentryALTinterwordspacing
T.~Van~Biemen, ``Personal privacy in practice: Putting the {GDPR} to test in a
  collective exercise of data subjects’ right of access,'' 2018. [Online].
  Available:
  \url{http://resolver.tudelft.nl/uuid:ccea2ec8-5ecb-47e3-8fae-79733d765093}
\BIBentrySTDinterwordspacing

\bibitem{kopieid}
``Hoe maak ik met de kopieid-app een veilige kopie van mijn
  identiteitsbewijs?''
  \url{https://www.rijksoverheid.nl/onderwerpen/identiteitsfraude/vraag-en-antwoord/veilige-kopie-
  identiteitsbewijs}.

\bibitem{uber_2021}
\BIBentryALTinterwordspacing
N.~Lomas, ``Dutch court rejects {Uber} drivers’ ‘robo-firing’ charge but
  tells {Ola} to explain algo-deductions,'' 2021-03-12. [Online]. Available:
  \url{https://techcrunch.com/2021/03/12/dutch-court-rejects-uber-drivers-robo-firing-charge-but-tells-ola-to-explain-algo-deductions/}
\BIBentrySTDinterwordspacing

\end{thebibliography}
\end{document}